\documentclass[preprint]{elsarticle}

\usepackage{graphicx}
\usepackage{lineno,hyperref}
\usepackage{amsmath,amsthm,amssymb}
\usepackage{url}
\usepackage{booktabs}
\usepackage{multirow}
\usepackage{ulem}
\usepackage{subfigure}
\usepackage{float}
\usepackage{color}
\usepackage[table,xcdraw]{xcolor}

\begin{document}

\begin{frontmatter}

\title{Modeling confirmation bias and peer pressure in opinion dynamics}

\author[1,2,4]{Longzhao Liu}
\author[1,4]{Xin Wang}
\author[1,4]{Shaoting Tang\corref{mycorrespondingauthor1}}
\cortext[mycorrespondingauthor1]{Corresponding author}
\ead{tangshaoting@buaa.edu.cn}
\author[1,3,4]{Zhiming Zheng}

\address[1]{LMIB, NLSDE, BDBC, School of Mathematical Sciences, Beihang University, Beijing 100191, China}
\address[2]{ShenYuan Honor School, Beihang University, Beijing 100191, China}
\address[3]{Institute of Artificial Intelligence and Blockchain, Guangzhou university, Guangdong province 510006, China}
\address[4]{PengCheng Laboratory, Shenzhen, 518055 , China}

\begin{abstract}
Confirmation bias and peer pressure both have substantial impacts on the formation of collective decisions. Nevertheless, few attempts have been made to study how the interplay between these two mechanisms affects public opinion evolution. Here we propose an agent-based model of opinion dynamics which incorporates the conjugate effect of confirmation bias (characterized by the population identity scope and individual stubbornness) and peer pressure (described by a susceptibility threshold). We show that the number of opinion fragments first increases and then decreases to one as population identity scope becomes larger in a homogeneous population. Further, in heterogeneous situations, we find that even a small fraction of impressionable individuals who are sensitive to peer pressure could help eliminate public polarization when population identity scope is relatively large. Intriguingly, we highlight the emergence of `impressionable moderates' who are easily influenced, hold wavering opinions, and are of vital significance in competitive campaigns.
\end{abstract}

\begin{keyword}
agent-based model, confirmation bias, peer pressure, public opinion dynamics
\end{keyword}

\end{frontmatter}

\section{Introduction}
People tend to accept claims that adhere to their prior beliefs, i.e., being within their identity scope, and ignore dissenting claims \cite{plous1993psychology,bessi2015science,zollo2017debunking}. This mechanism, confirmation bias, causes the emergence of echo chambers and group polarizations in public discourse \cite{nickerson1998confirmation,baumann2020modeling}. Political polarization is a prominent case where Republicans reject the statements supporting Democrats as false and vice versa, which threatens the democracy \cite{bail2018exposure}. In addition, individuals are likely to reshape their opinions, attitudes or behaviors according to the position of the majority \cite{rogers2010diffusion,granovetter1978threshold}. This mechanism, peer pressure, appears to be a primary driver of opinion evolution \cite{centola2010spread,he2020opinion}. It is sometimes strategically utilized by partisan organizations to get more votes in election, for instance, deploying vast social bots \cite{bessi2016social} and information gerrymandering \cite{stewart2019information}. Thus, to comprehensively understand the opinion polarization and to forecast the winner in competing processes require a model of opinion dynamics integrating the conjugate effect of confirmation bias and peer pressure.

The development of online networks has radically changed the way that people consume information and exchange opinions, which results in substantial impacts on opinion dynamics \cite{kleinberg2013analysis,lewis2012social,schmidt2017anatomy,wang2017promoting}. Individuals could easily seek out contents coherent with prior beliefs on large-scale online discourse, which amplifies confirmation bias \cite{quattrociocchi2011opinions}. In addition, the wide availability of dissenting content on the web makes individuals exposed to peer pressure every day \cite{jang2016social}. Under such circumstance, opinion dynamics has attracted great attentions recently \cite{holme2006nonequilibrium}. Classical models including Friedkin and Johnsen model \cite{friedkin2016network,friedkin2006structural,parsegov2016novel}, Sznajd model \cite{sznajd2000opinion} and voter model \cite{klamser2017zealotry} were further explored on large-scale social networks. These models sufficiently considered interpersonal influence including peer pressure between discordant pairs and showed the consensus state where all agents finally share the same opinion.

In addition to group consensus, empirical studies showed fragmentation and polarization of opinions, which has aroused great concerns in diverse fields \cite{axelrod1997dissemination,liu2020homogeneity,quattrociocchi2016echo}. For instance, political scientists observed markedly increased political polarization in the United States, which threatens democracies \cite{crockett2017moral}. Sociologists noted the aggregation of individuals trusting false news and misinformation which have been recorded as one of main threats to human society by World Economic Forum \cite{del2016spreading,howell2013digital}. Various classical mathematical models were proposed to explain the ubiquitous phenomena \cite{castellano2009statistical,shao2009dynamic,gleeson2014competition}. Deffuant {\it et al.} proposed bounded confidence model (BCM) where individuals just accept contents within their identity scope to mimic confirmation bias and found the occurrence of several opinion clusters when identity scope is small \cite{deffuant2000mixing}. Vicario {\it et al.} incorporated negative updating rule of opinions among discordant pairs of users into BCM and observed the coexistence of two stable final opinions \cite{del2017modeling}. Wang {\it et al.} introduced an agent-based model integrating external political campaigns and opinion dynamics, which reproduced the 2016 American presidential election well and yielded the intriguing moderate clusters in addition to two polarized clusters \cite{wang2020public}.

While confirmation bias and peer pressure are not necessarily new phenomena, few attempts have been made to explore how the interplay between the two mechanisms affects opinion evolution. Besides, moderate cluster but not polarized clusters usually determines the winner in competing processes of opinions. Presidential election is a prominent case where the party who wins the majority of moderates would win. Unveiling the facets of moderates is a meaningful problem which is of vital significance to design efficient strategies of guiding public opinions.

In this paper, we propose a modified bounded confidence model with the conjugate effect of confirmation bias and peer pressure to describe the continuous opinion dynamics on large-scale social networks. Firstly, we consider population with homogeneous susceptibility to peer pressure. Simulations results show intriguing non-monotonous changes in the number of opinion fragments, which first increases and then reduces to unique one with identity scope increasing. Then we consider heterogeneous population where impressionable individuals who are sensitive to peer pressure and 'confident' individuals who are immune to peer pressure coexist. Results show critical phenomena about the critical fraction of impressionable individuals leading to public consensus: the value is very small when identity scope is relatively large, while explosively increases to a large one once identity scope decreases to a threshold. In addition, we highlight that majority of impressionable individuals would become moderates. It indicates the emergence of an important but easily influenced group: impressionable moderates. This result implies the insight that targeting impressionable moderates might be an efficient strategy to guide public opinions even when the system has reached steady state.

\section{Model}
\begin{figure}
\center
\includegraphics[width=12cm]{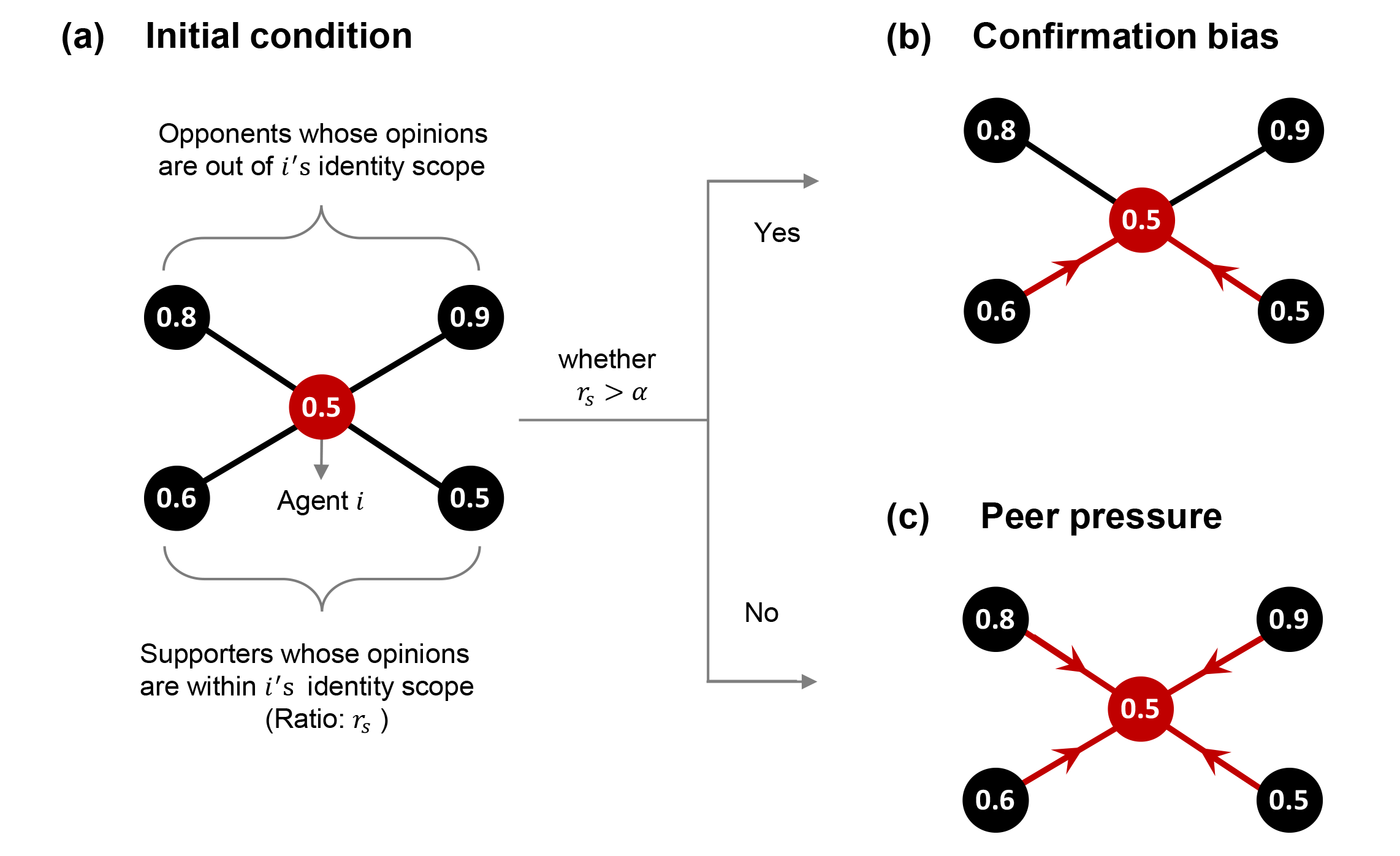}

\caption{Schematic of dynamical model. (a) Initial condition. Agent $i$ passively receives information from neighbors. According to their claims, agent $i$ divides neighbors into two classes: supporters whose opinions are within its identity scope (nodes in the bottom) and opponents whose opinions are out of its identity scope (nodes in the top). If the ratio of supporters is more than a threshold ($r_s>\alpha$), agent $i$ would go through (b) confirmation bias and tends to update its opinion in its comfortable cognitive region by only interacting with supporters. Otherwise, agent $i$ experiences (c) peer pressure and updates its opinion by interacting with all neighbors.}
\label{model}
\end{figure}

We introduce an agent-based model of continuous opinion dynamics which integrates the conjugate effect of confirmation bias and peer pressure in this section. Consider a large-scale social network with $N$ agents. Initially, each agent has an opinion $x_i$ that is uniformly distributed in [0,1]. At each time step, social platforms allow agents to rapidly receive opinions from all their neighbors. According to the well-known confirmation bias, agents just think that opinions close enough to their beliefs are reasonable and that corresponding neighbors are supporters of their cognition. Complying with previous studies \cite{wang2020public}, we define $\delta$ as identity scope to describe the phenomena. Specifically, neighbor $j$ is recognized as a supporter of agent $i$ if $|x_i-x_j|<\delta$. Otherwise, neighbor $j$ is recognized as an opponent. In this way, any agent could divide its neighbors into two classes: supporters and opponents (see figure \ref{model}(a)). Another important mechanism in opinion dynamics is peer pressure that individuals are likely to reshape their cognition, attitudes or behaviors when being affected by others, especially by opponents \cite{haun2011conformity,stevens1996classification}. Here we adopt the thought of threshold model to mimic peer pressure, where we define threshold $\alpha$ as susceptibility to peer pressure. If the fraction of supporters is more than the threshold $\alpha$, agents would affirm their cognition and readjust opinions by just interacting with neighbor supporters, which is shown as confirmation bias in figure \ref{model}(b). Otherwise, agents would doubt their own cognition due to lack of support and heavy peer pressure from opponents, which urges them to update opinions according to all neighbors' position (see figure \ref{model}(c)).

Specifically, our dynamical model goes through the following steps:
\begin{itemize}
\item[1)] At each time step, agent $i$ divides its neighbors into supporters and opponents, according to the following rule: if $|x_i-x_j|<\delta$, neighbor $j$ is recognized as a supporter. Otherwise, $j$ is an opponent.
\item[2)] At each time step, one of the two possible interactions would happen.
\begin{itemize}
\item [(a)]({\it Confirmation bias}) If the fraction of supporters is more than $\alpha$, agent $i$ updates its opinion from $x_i$ to $\tilde{x_i}$ by interacting with supporters: $$\tilde{x_i}=\mu x_i+(1-\mu) \sum\limits_{j\in \Omega_i^S}\frac{x_j}{|\Omega_i^S|}$$
    where $\mu$ reflects the stubbornness and $\Omega_i^S$ represents the set of neighbor supporters of agent $i$.
\item [(b)]({\it Peer pressure}) Otherwise, agent $i$ updates its opinion from $x_i$ to $\tilde{x_i}$ by interacting with all neighbors:
    $$\tilde{x_i}=\mu x_i+(1-\mu) \sum\limits_{j\in \Omega_i}\frac{x_j}{|\Omega_i|}$$
    where $\Omega_i$ represents the set of neighbors of agent $i$.
\end{itemize}
\end{itemize}

\section{Results}
\subsection{Homogeneous population with uniform susceptibility to peer pressure}
Here we perform simulations of our model on large-scale Erd\"{o}s-R\'{e}nyi (ER) network with $N=50000$ nodes, whose average degree is $\langle k\rangle=40$. Each simulation runs for 500 steps to ensure that the dynamical system reaches the stable state. We divide the opinion interval [0,1] into 100 bins and compute the frequency of opinion values falling into each bin. Complying with previous studies \cite{del2017modeling}, we utilize the number of peaks in distribution of opinion frequency to represent the number of final opinion fragments, where two peaks are regarded as separate if the distance between them is more than 0.1.

\begin{figure}
\center
\includegraphics[width=12cm]{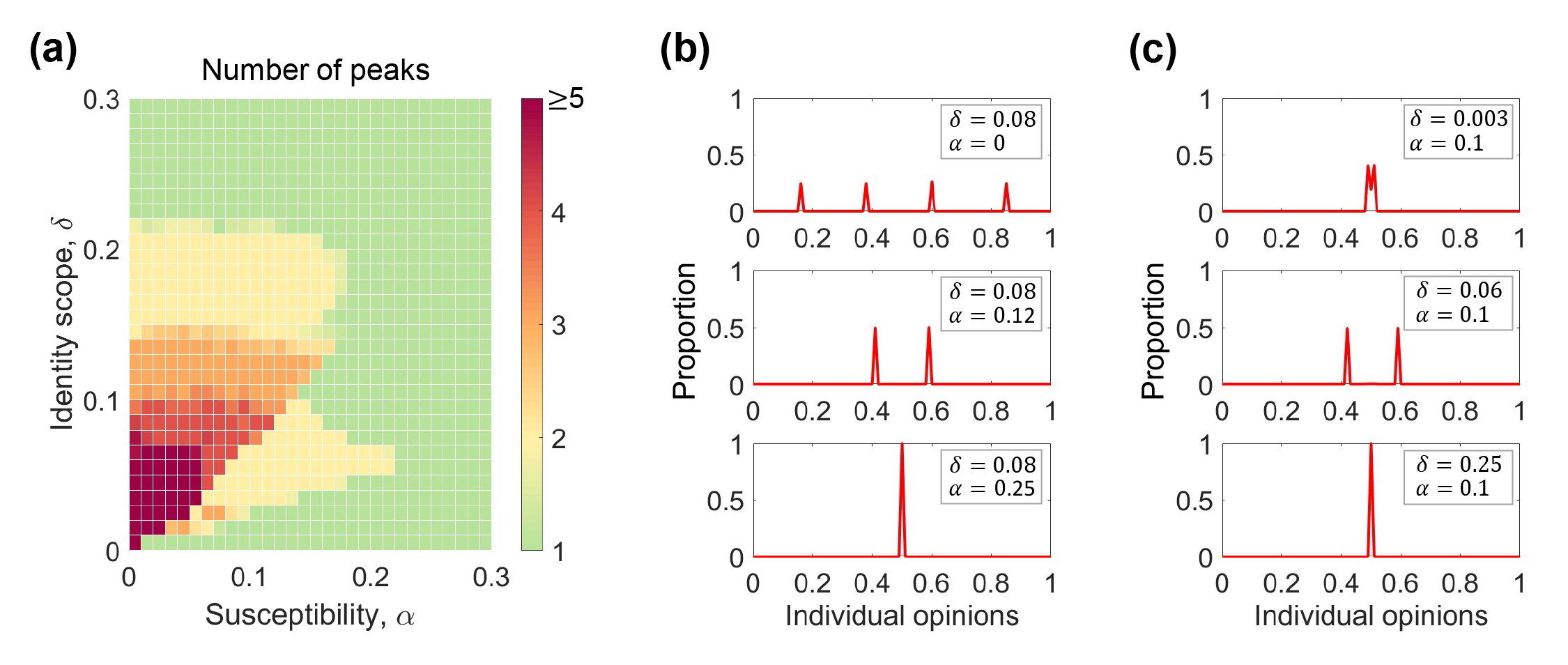}

\caption{Non-monotonous conjugate effect of confirmation bias and peer pressure. (a) Shown is the number of peaks in distribution of opinion frequency under different combinations of identity scope ($\delta$) and susceptibility to peer pressure ($\alpha$). The number of peaks decreases with $\alpha$ growing, which is again confirmed by showing the detailed distribution of cases with varying $\alpha$ in (b). Moreover, we observe non-monotonous conjugate effect when $\alpha\neq 0$: the number of peaks first increases and then reduces to one with $\delta$ growing. The details of some cases are shown in (c). Parameters: $\mu=0.3$.}
\label{non-monotonous effect}
\end{figure}

In figure \ref{non-monotonous effect}, we explore how the interplay between confirmation bias and peer pressure affects the distribution of final opinions. Figure \ref{non-monotonous effect}(a) presents phase diagram for the number of peaks in opinion distribution under different combinations of identity scope ($\delta$) and susceptibility to peer pressure ($\alpha$). All simulation results are averaged over 5 independent runs and the robustness is shown in figure \ref{robustness} (\ref{appendix1}). In the phase diagram, we observe decrease of peak number with respect to increase in susceptibility to peer pressure, which is again confirmed by showing the detailed distribution in figure \ref{non-monotonous effect}(b). The result adheres to our intuition that peer pressure promotes the consensus of public opinions. More interestingly, we highlight that the number of peaks first increases and then reduces to one with identity scope $\delta$ growing when $\alpha\neq 0$. The non-monotonous changes are illustrated by showing detailed distribution in figure \ref{non-monotonous effect}(c), which can not be observed in classical Bounded Confidence Model \cite{deffuant2000mixing}. While the increase of peak number with $\delta$ growing is counter-intuitive, it can be explained by the following intuitive reasons. In the same surroundings, individuals with smaller identity scope think that more opinions deviate from their beliefs, and thus experience more peer pressure, which urges them to make a major change in their opinions in order to adapt to the public.

\begin{figure}
\center
\includegraphics[width=12cm]{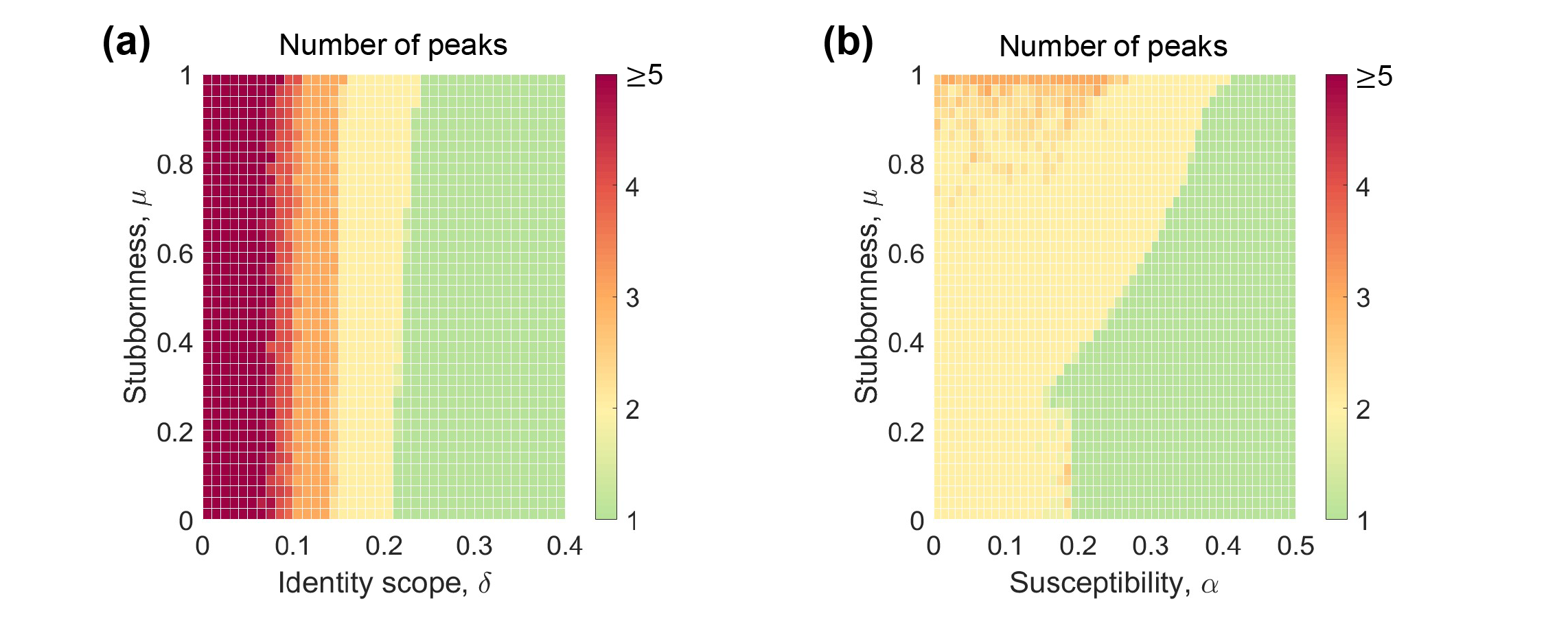}

\caption{How individual stubbornness affects the final opinions. Phase diagrams for the number of peaks in opinion distribution are shown (a) under different combinations of stubbornness and identity scope, and (b) under different combinations of stubbornness and susceptibility to peer pressure. Simulation results are averaged over 5 independent runs. We find that critical value of $\delta$ leading to public consensus shows almost no change with stubbornness $\mu$ growing, while critical value of $\alpha$ increases significantly. Parameters: (a) $\alpha=0$; (b) $\delta=0.15$.}
\label{Stubbornness}
\end{figure}

Through observing the phase diagram in figure \ref{non-monotonous effect}, we find that public consensus could be caused by two conditions: large identity scope ($\delta$) or large susceptibility to peer pressure ($\alpha$). Here we focus on how individual stubbornness affects critical conditions for public consensus. Figure \ref{Stubbornness} presents the number of peaks in final opinion distribution under different combinations of stubbornness and identity scope (figure \ref{Stubbornness}(a)) or susceptibility to peer pressure (figure \ref{Stubbornness}(b)). Result shows almost no change in critical value of $\delta$ with stubbornness varying, which is consistent with insights provided by Bounded Confidence Model. Different from the tiny effect on critical value of $\delta$, we find that individual stubbornness significantly influences the critical value of $\alpha$.

\subsection{Heterogeneous population with different susceptibilities to peer pressure}
Note that the susceptibility to peer pressure in population is heterogeneous. We consider the scenario where the population consists of two classes: impressionable individuals who would be affected by peer pressure if less than half the neighbors support them and 'confident' individuals who are completely immune to peer pressure. The values of $\alpha$ corresponding to these two classes are $0.5$ and $0$, respectively.  We denote the fraction of impressionable individuals as $\rho$.

\begin{figure}
\center
\includegraphics[width=12cm]{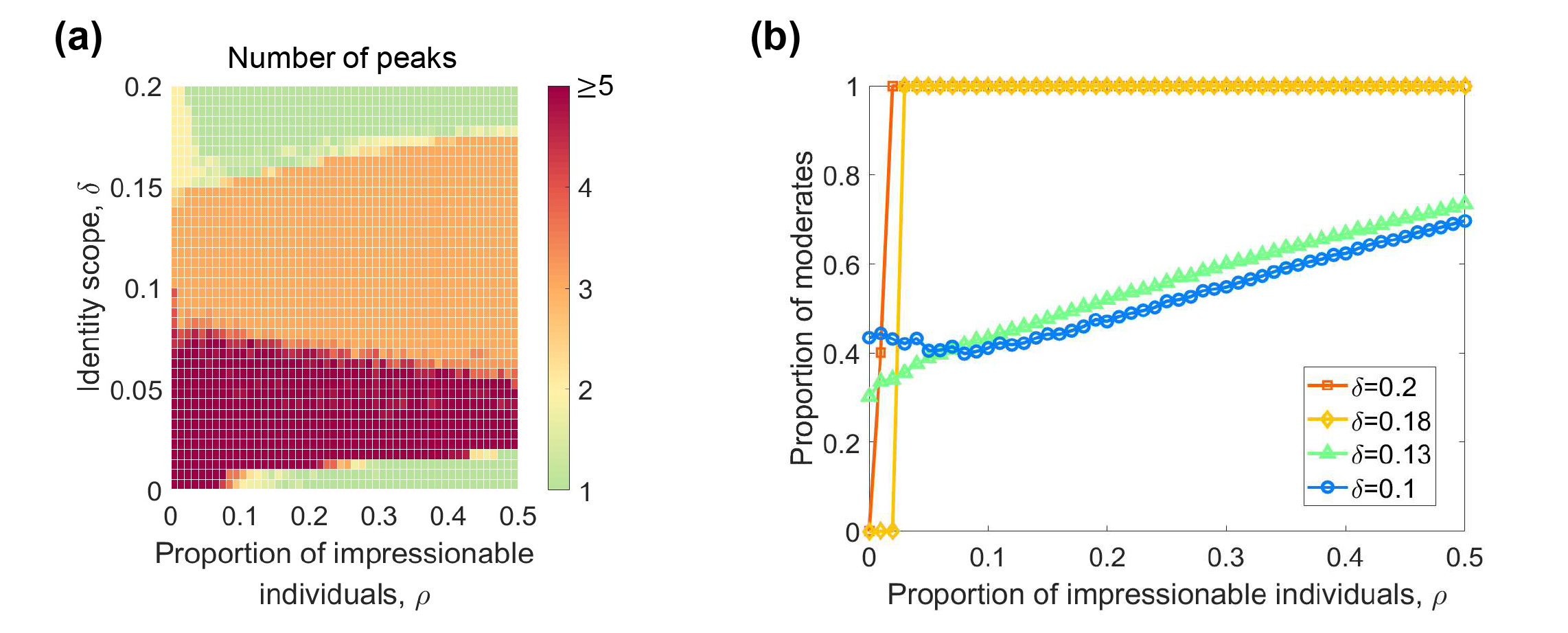}

\caption{Effect of impressionable individuals. (a) How impressionable individuals affect the number of peaks in final opinion distribution. Shown is phase diagram under different combinations of identity scope ($\delta$) and the fraction of impressionable individuals ($\rho$). Small $\rho$ is enough for the public to reach consensus when $\delta$ is relatively large, while the increase of $\rho$ does not reduce peak number when $\delta$ is relatively small. (b) How impressionable individuals affect the proportion of moderates at steady state. Shown is the proportion of moderates with respect to changes in $\rho$. Parameters: $\mu=0.3$.}
\label{impressionable individuals}
\end{figure}

Firstly, we explore how impressionable individuals affect opinion evolution in public discourse. Figure \ref{impressionable individuals}(a) shows the number of peaks in opinion distribution under different combinations of identity scope ($\delta$) and fraction of impressionable individuals ($\rho$). Results show the intriguing critical phenomena when considering critical value of $\rho$ leading to public consensus as a function of $\delta$: critical value of $\rho$ explosively increases as $\delta$ decreases to 0.16. To be specific, small $\rho$, no more than 0.05, could lead to public consensus when $\delta$ is relatively large ($\delta\in [0.16,0.2]$), while very large $\rho$, equal to 0.5, is not enough for public consensus when $\delta$ is relatively small ($\delta< 0.16$). Thus, the premise that a few impressionable individuals could efficiently eliminate public polarization is relatively large identity scope in population.

In addition, we observe the existence of three stable opinion fragments in large region of figure \ref{impressionable individuals}(a), which occurs in many real-world scenarios such as presidential election \cite{wang2020public}. In particular, moderate group is of vital significance as their choices directly determine the winner in competing processes. This motivates us to explore the characteristics of the moderate group generated by our dynamical model. Here moderates refer to individuals whose opinions belong to $[0.5-V, 0.5+V]$, where $V$ is a parameter. It is proven that the value of $V (V\in[0.01,01])$ has little impact on the proportion of moderates (figure \ref{V} in \ref{appendix2}). Here we set $V$ as 0.05.

Then we present the fraction of moderates as a function of $\rho$ under different identity scope in figure \ref{impressionable individuals}(b). For the case $\delta=0.2$ and $\delta=0.18$, the explosive increase of  moderates from 0 to 1 again illustrates that even a small number of impressionable individuals could eliminate the public polarization when $\delta$ is relatively large. For the case $\delta=0.13$ and $\delta=0.1$, the fraction of moderates slowly increases with impressionable individuals increasing. These indicate the positive relationship between impressionable individuals and moderates.

\begin{figure}
\center
\includegraphics[width=12cm]{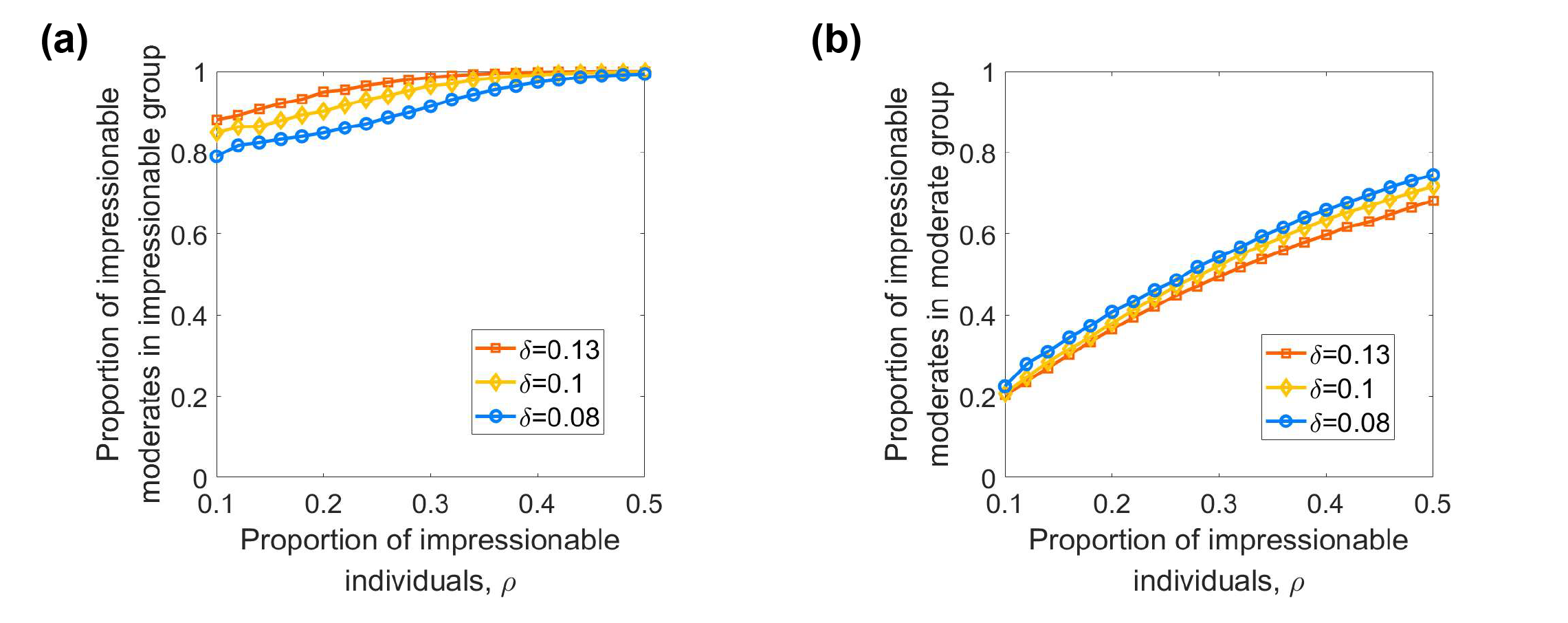}

\caption{Emergence of vast impressionable moderates. Shown is the fraction of impressionable moderates (a) in all impressionable individuals and (b) in moderate group, with respect to changes in $\rho$. More than 80\% impressionable individuals finally evolve to moderates. Besides, the proportion of impressionable moderates in moderate group increases with $\rho$ growing. Parameters: $\mu=0.3$.}
\label{moderates}
\end{figure}

Furthermore, we focus on a special group: impressionable moderates. The group is more likely to be affected when experiencing external peer pressure but their decisions play a determinant role in competing processes. We present the fraction of impressionable moderates in impressionable group (figure \ref{moderates}(a)) and in moderate group (figure \ref{moderates}(b)), respectively. Results highlight that while initial opinions of impressionable individuals are randomly distributed, more than 80\% of them finally become moderates in most scenarios. In addition, the fraction of impressionable moderates in moderate group continuously increases with $\rho$ growing. These indicate the emergence of vast impressionable moderates at the steady state of systems where $\rho$ is relatively large. It naturally inspires us a insight that targeting impressionable moderates might be an efficient strategy to guide public opinion even when the system has reached equilibrium. A typical example is presidential election where deploying numerous zealots gives impressionable moderates much peer pressure to guide their opinions.

\section{Conclusions and discussions}
Opinion evolution on large-scale social networks has been widely concerned \cite{hegselmann2002opinion,acemouglu2013opinion,qian2011adaptive}. People are more likely to accept claims within their identity scope and ignore the dissenting claims \cite{zhu2010individual,frenda2011current}. On the contrary, people might alter their decisions if receiving vast opposing views \cite{harden2008gene}. Both confirmation bias and peer pressure were proved as core factors in opinion dynamics, separately \cite{sunstein1999law,watts2002simple}. However, it remains unclear how the interplay between these two mechanisms affect opinion evolution.

In this work, we propose an agent-based model of opinion evolution which takes into account both confirmation bias and peer pressure. By performing simulations in homogeneous population, we find the non-monotonous conjugate effect of these two mechanisms. To be specific, the number of opinion fragments first increases and then reduced to one with identity scope growing, which is completely different from the insight of classical bounded confidence model \cite{deffuant2000mixing}. The counter-intuitive phenomena can be explained by the following microscopic reason: agents with smaller identity scope consider that more opinions deviate from their beliefs and thus receive more peer pressure which urges them to adapt to the public.

In addition, we consider the heterogeneity of susceptibility to peer pressure. Specially, the heterogeneous population is divided into impressionable individuals and individuals immune to peer pressure. We explore the detailed effect of impressionable individuals. First, we find intriguing critical phenomena that critical fraction of impressionable individuals resulting in public consensus explosively increases from a tiny value to a large value when identity scope goes through a threshold. This result indicates that only when identity scope is relatively large can a few impressionable individuals efficiently eliminate polarization of public. Moreover, we highlight that while impressionable individuals' initial opinions are randomly distributed, more than 80\% of them finally become moderates when system reaches the steady state. It reflects an important but easily influenced group: impressionable moderates.

Our work utilizes simple mechanisms to reveal the non-monotonous conjugate effect of confirmation bias and peer pressure and explains the counter-intuitive phenomena from microscopic level. Besides, the study about heterogeneity of susceptibility to peer pressure shows a powerful method of eliminating the polarization of public when identity scope is relatively large, which is deploying a few impressionable individuals. In real online networks, we could deploy a small number of social bots to replace impressionable individuals. Moreover, our model shows the emergence of impressionable moderates. The important but easily influenced group would naturally become targets of interest to all parties in competing processes. The strategies of targeting impressionable individuals deserves further study.

\section*{Acknowledgement}
This work is supported by Program of National Natural Science Foundation of China Grant No. 11871004, 11922102, and National Key Research and Development Program of China Grant No. 2018AAA0101100.

\begin{appendix}
\section{Robustness of simulation results} \label{appendix1}
\begin{figure}[p]
\center
\includegraphics[width=12cm]{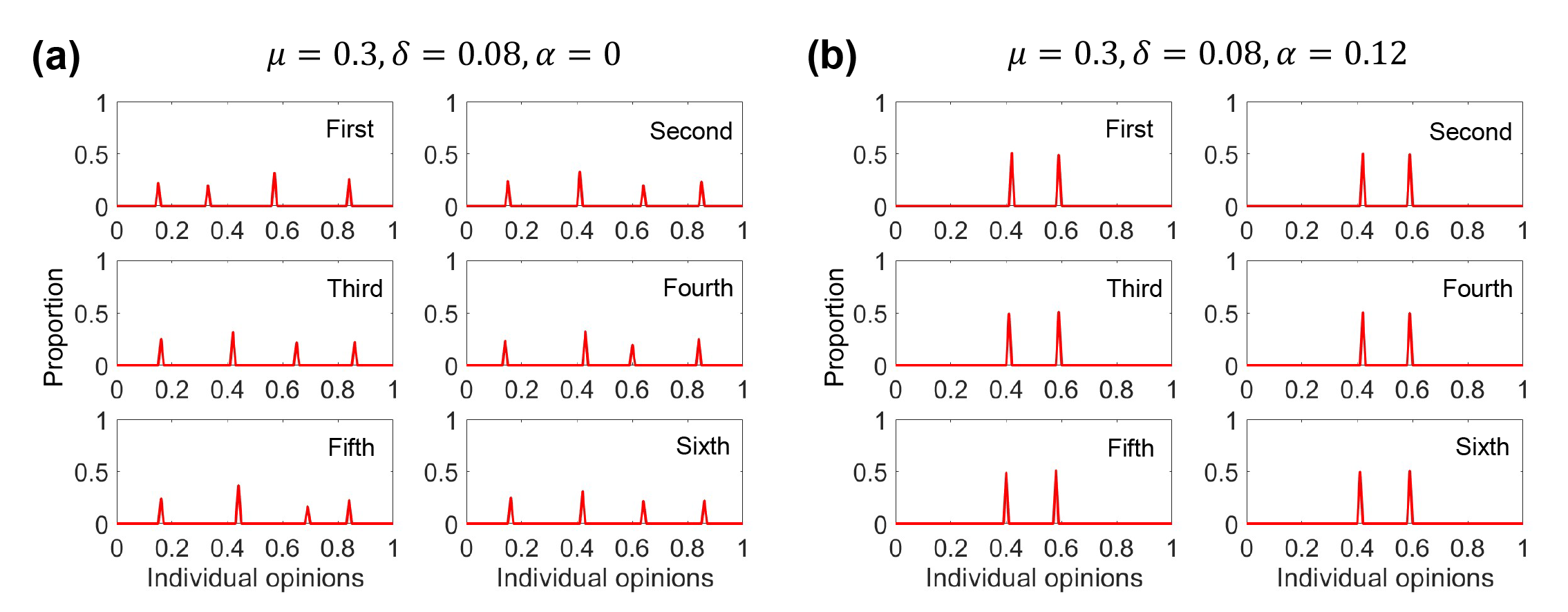}

\caption{Robustness of simulation results. For each combination of parameters, we perform simulations of our model for six independent times. Distribution of final opinions in each simulation is shown in subplot. Parameters: (a) $\mu=0.3$, $\delta=0.08$, $\alpha=0$; (b) $\mu=0.3$, $\delta=0.08$, $\alpha=0.12$.}
\label{robustness}
\end{figure}

Figure \ref{robustness} presents the distribution of final stable opinions, where each subplot corresponds to one independent simulation under the same combination of parameters. Note that these subplots under the same combination are similar in figure \ref{robustness}(a) or figure \ref{robustness}(b). It indicates that the simulation results about the number of peaks in opinion distribution are robust.

\section{Robustness of results under different values of $V$} \label{appendix2}
\begin{figure}[p]
\center
\includegraphics[width=7cm]{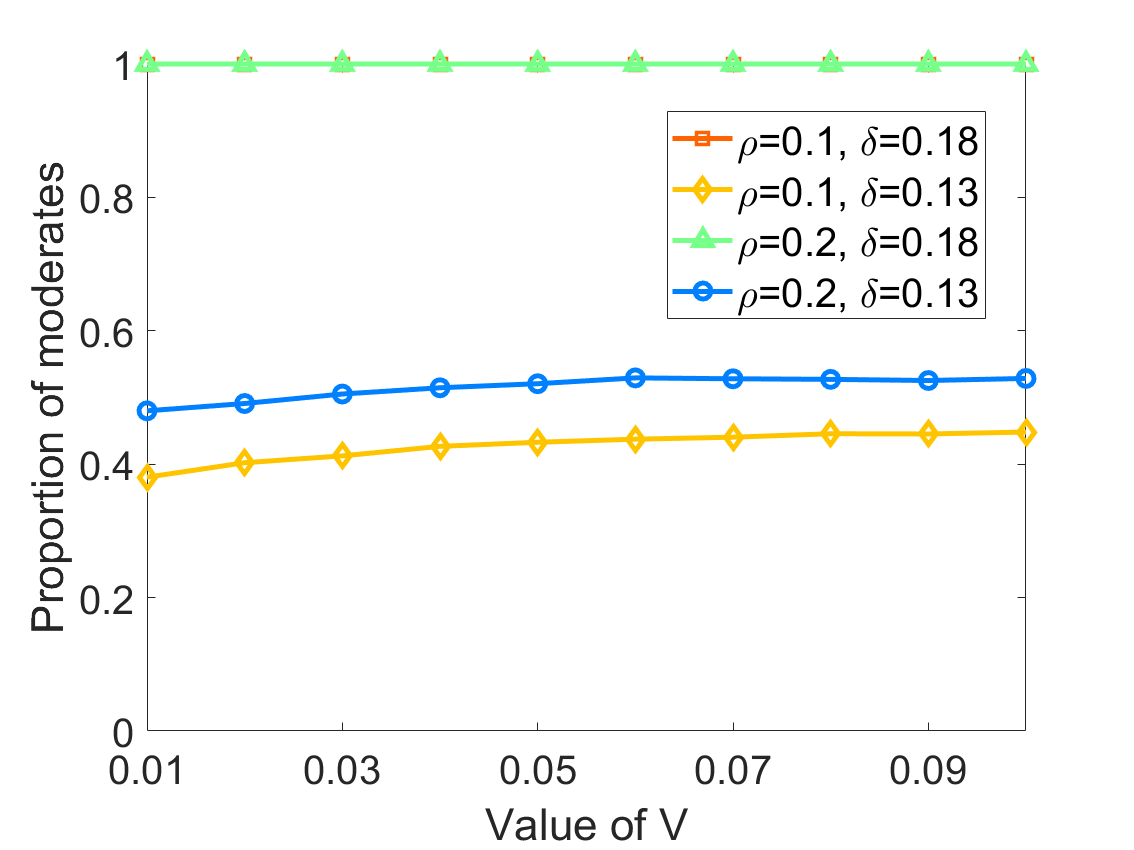}

\caption{Robustness of results under different values of $V$. Shown is the proportion of moderates as a function of $V$ under different combinations of parameters. Results show that proportion of moderates is almost the same with $V$ varying. Parameters: $\mu=0.3$.}
\label{V}
\end{figure}

The value of parameter $V$ determines the opinion boundary of moderates. Figure \ref{V} presents the proportion of moderates with respect changes in parameter $V$. We find that the value of $V$ ranging from 0.01 to 0.1 almost shows no influence on the proportion of moderates.

\end{appendix}


\end{document}